\documentclass[nofootinbib,letterpaper,preprint,preprintnumbers,amsmath,amssymb]{revtex4}
\usepackage{latexsym}
\usepackage{hyperref}
\begin{document}
\title{\bf Einstein Product Metrics in Diverse Dimensions}
\author{K. R. Koehler}
\affiliation{University of Cincinnati / Raymond Walters College, Cincinnati, OH 45236}
\email{kenneth.koehler@uc.edu}
\begin{abstract}
We use direct products of Einstein Metrics to construct new solutions to Einstein's Equations with 
cosmological constant. We illustrate the technique with three families of solutions having the geometries
Kerr/de Sitter $\otimes$ de Sitter, Kerr/anti-de Sitter $\otimes$ anti-de Sitter and Kerr $\otimes$ Kerr.
\end{abstract} \maketitle
Shortly after Einstein introduced the cosmological constant to General Relativity, Kasner \cite{1} and
Schouten and Struik \cite{2} explored the geometry of Einstein Manifolds: manifolds admitting a metric
whose Ricci Tensor is a constant multiple of the metric. These are manifolds of constant curvature and
solutions to Einstein's Equations with zero stress-energy tensor but possibly nonzero
cosmological constant.
With Hubble's discovery in 1929 of the expanding universe, Einstein's motivation for introducing the cosmological 
constant was lost and the canonical Einstein Manifolds, de Sitter and anti-de Sitter spacetimes,
were largely relegated to serve as textbook examples. Then in 1998, measurements of
high red-shift Type Ia supernovae \cite{3} indicated that the expansion of the universe was accelerating,
possibly due to a positive cosmological constant.

Interest in manifolds with dimension greater than 4 has paralleled that of interest in the cosmological constant.
As early as 1914 \cite{4}, a fifth dimension was postulated as a mechanism for unifying the electromagnetic
and gravitational forces. Kaluza and Klien independently applied this mechanism in the context of
General Relativity, but for nearly half a century it was seen as unphysical for reasons which are still
obvious. Since then, however, the progress in understanding supergravity and string theories has carried
interest in manifolds of diverse dimension throughout the academic community and beyond into the public
consciousness.

In this note we construct direct products of Einstein Metrics which are solutions to Einstein's
Equations with cosmological constant in diverse dimensions. For two or more metrics defined on disjoint
manifolds, the product manifold carries a metric which is the simple sum
\begin{eqnarray}
ds^2 = ds_1^2(x) + ds_2^2(y) + \ldots
\end{eqnarray}
For such metrics, the Christoffel Connection and Riemann and Ricci Tensor components
are simple sums of the corresponding components on the submanifolds. In addition,
the scalar curvature and the Kretschmann Invariant are simple sums of the corresponding
invariants. The utility of these relationships has been known and exploited for over
half a century to simplify manual computations of connection and tensor components.
Using them we see that for the product metric, Einstein's Equations \emph{almost} uncouple into
a set of disjoint equations:
\begin{eqnarray}
R_{i, a b} + (\Lambda - \frac{1}{2}
\sum_j R_j ) 
g_{i, a b} = \alpha T_{i, a b}
\end{eqnarray}
It is clear that if the component metrics are independently solutions to Einstein's Equations with 
vanishing scalar curvatures, the 
product metric is also a solution. This implies, for instance, that any product of 
Ricci-flat metrics (ie., circles, flat tori, Minkowski and Kerr spacetimes, or their Euclidean counterparts)
is a solution.

If the component metrics are Einstein Metrics, we have
\begin{eqnarray}
R_{i, a b} &=& \chi_i g_{i, a b}
\cr
R_i &=& D_i \chi_i
\cr
\Lambda_i &=& \frac{D_i - 2}{2} \chi_i
\end{eqnarray}
where $\chi_i$ is a constant and $D_i$ is the dimension of the 
$i^{th}$ metric. Note that for $D_i < 3$, $\Lambda_i = 0$.

For a product of $n$ Einstein Metrics $g_i$, Einstein's Equations reduce to a set of $n$ algebraic relations
\begin{eqnarray}
\chi_i + \Lambda - \frac{1}{2}
\sum_j R_j = 0.
\end{eqnarray}
Obviously the $\chi_i$ must all be equal for the product metric to be
a solution, which fixes
$\Lambda$ and the $\Lambda_i$ in terms of 
$\chi$:
\begin{eqnarray}
\Lambda &=& \chi \frac{\sum_j D_j - 2}{2}
\cr
\Lambda_i &=& \frac{D_i - 2}{2} \chi
\end{eqnarray}
It is clear that the cosmological constants must all have the same sign.

The most commonly discussed Einstein Manifolds which are not Ricci-flat are de Sitter,
anti-de Sitter and Kerr with nonzero cosmological constant. A 
Euclidean de Sitter (spherical) metric for $D_S > 2$ is \cite{5}
\begin{eqnarray}
ds_S^2 = \beta^2 
\cosh^2 \frac{t}{\beta}
d\Omega_{D_S - 1}^2 + dt^2
\end{eqnarray}
where $d\Omega_i^2$ is the standard metric on $S^i$.
For either this metric or its Lorentzian counterpart (or for the standard metric on $S^2,
ds_S^2 = \beta^2 
d\Omega_2^2$),
we find
\begin{eqnarray}
\chi_S = \frac{D_S - 1}{\beta^2}
\end{eqnarray}

A Euclidean anti-de Sitter metric for $D_A > 2$ is \cite{5}
\begin{eqnarray}
ds_A^2 = \alpha^2 (dr^2 + \sinh^2 r
d\Omega_{n-2}^2 + \cosh^2 r dt^2)
\end{eqnarray}
We include the constant factor $\alpha$ which in the product metric
becomes a nontrivial parameter. For either this metric or its Lorentzian counterpart,
\begin{eqnarray}
\chi_A = - \frac{D_A - 1}{\alpha^2}
\end{eqnarray}

Kerr solutions with nonzero cosmological constant \cite{6} exist in any dimension $D_K > 3$. Let
\begin{eqnarray}
\rho^2 &=& r^2 + a^2 \cos^2 \theta
\cr
\Delta(D_K) &=& (1 - \frac{2 \Lambda_K r^2}{(D_K - 1) (D_K - 2)}) (r^2 + a^2) - \frac{\mu}{r^{D_K-5}} 
\cr
\psi(D_K) &=& 1 + \frac{2 a^2 \Lambda_K \cos^2 \theta}{(D_K - 1) (D_K - 2)}
\cr
\Sigma(D_K) &=& 1 + \frac{2 a^2 \Lambda_K}{(D_K - 1) (D_K - 2)}
\end{eqnarray}
where $\mu$ is proportional to the mass, $a$ is proportional to the angular momentum
and $\Lambda_K$ is the
cosmological constant. The corresponding Kerr metric is
\begin{eqnarray}
ds_K^2 &=& r^2 \cos^2 \theta 
d\Omega^2 +
\frac{\rho^2}{\Delta(D_K)} dr^2 + 
\frac{\rho^2}{\psi(D_K)} d\theta^2 + \cr
&&\frac{(\psi(D_K) (r^2 + a^2)^2 - 
\Delta(D_K)
a^2 \sin^2 \theta)
\sin^2 \theta}{\Sigma^2(D_K)
\rho^2} d\phi^2 + \cr
&&\frac{2 a \sin^2 \theta (\psi(D_K) 
(r^2 + a^2) -
\Delta(D_K))}{\Sigma^2(D_K)
\rho^2} d\phi dt - \cr
&&\frac{\Delta(D_K) - a^2 \psi(D_K) 
\sin^2 \theta}{\Sigma^2(D_K)
\rho^2} dt^2
\end{eqnarray}
where $d\Omega^2$ is the standard metric on $S^{D_K-4}$.
These solutions possess a single rotation axis; more general solutions have been discussed
\cite{7} which have the maximal number of rotation parameters. For these metrics or their Euclidean counterparts,
\begin{eqnarray}
\chi_K = \frac{2 \Lambda_K}{D_K - 2}
\end{eqnarray}

Forming the product metric $ds^2 = ds_K^2 + ds_S^2$, we
set $\chi_K$ equal to $\chi_S$ and find
the product is a solution for
\begin{eqnarray}
\Lambda &=& \frac{(D_K + D_S - 2) \Lambda_K}{D_K - 2}
\cr
\beta &=& \sqrt{\frac{(D_K - 2) (D_S - 1)}{2 \Lambda_K}}
\end{eqnarray}

Physically, an observer on the Kerr submanifold measures a smaller value for the cosmological constant
than an observer living in the entire product spacetime, and the ratio of the two is purely a function of
the dimension of the spherical submanifold. In addition, the cosmological constant 
measured by the Kerr observer is inversely proportional to the square of the radius parameter for the sphere.
Assuming that $D_K = 4$ and $\Lambda_K$ corresponds to
data from current observations \cite{8}, we find that $\beta$ is of the order of the radius of 
the observable universe. Since the geodesic equations for a simple product manifold are not coupled,
each submanifold is a geodesic hypersurface. But without a mechanism to prevent energy and momentum
transfer between submanifolds, these manifolds are clearly of purely academic interest.

This is obviously true as well when more than one submanifold is noncompact. For the anti-de Sitter case,
$ds^2 = ds_K^2 + ds_A^2$ is a solution if
\begin{eqnarray}
\Lambda &=& \frac{(D_K + D_A - 2) \Lambda_K}{D_K - 2}
\cr
\alpha &=& \sqrt{\frac{- (D_A - 1) (D_K - 2)}{2 \Lambda_K}}
\end{eqnarray}

Finally, we form a product of two Kerr metrics with cosmological constant and confirm that the solution
requires
\begin{eqnarray}
\Lambda &=& \frac{(D_1 + D_2 - 2) \Lambda_1}{D_1 - 2}
\cr
\Lambda_2 &=& \frac{(D_2 - 2) \Lambda_1}{D_1 - 2}
\end{eqnarray}

We noted previously that the scalar curvature and Kretschmann Invariant for a simple product metric
are simply the sums of those invariants for the metrics on the submanifolds.
Therefore if either metric possesses curvature singularities, the product metric possesses those same singularities.
This means that our example product metrics possess the well-known ring singularity for $D_K (D_i)$
equal to 4 or 5, and the singularity at $r = 0$ for $D_K (D_i) > 5$ \cite{9}.)

\section*{Acknowledgements}
We would like to thank Cenalo Vaz and Louis Witten for stimulating discussions on these matters.

\end{document}